# HIGH-SPIN TERMINATING BANDS FORMED FROM SUCCESSIVE PARTICLE-HOLE EXCITATIONS ACROSS CLOSED SHELLS*

INGEMAR RAGNARSSON

Div. of Mathematical Physics, Lund Inst. of Technology, P.O. Box 118, S-221 00 Lund, Sweden

AND

W.C. MA

Dept. of Physics, Mississippi State University, Mississippi State, MS 39762, USA

High spin states in $_{30}$Zn and $_{66}$Dy isotopes with two protons outside the closed shells at $Z = 28$ and $Z = 64$, respectively, are discussed. Bands terminating at higher and higher spin values are formed from successive proton excitations across these shells. These bands show very different features depending on the number of excited particles. Very good agreement with experiment is obtained. Special emphasis is put on the formation of $M1$-bands in configurations with one hole in the $Z = 64$ core.

## 1. Introduction

An important property of normal-deformed nuclear configurations is that their spin content is limited. Thus, they can generally be described in terms of a closed core + a number of valence particles (and/or holes). While the core cannot give any contribution to the angular momentum, the valence particles, which can be described as belonging to one or several $j$-shells, have a maximum spin value, $I_{max}$ which is easily calculated. Except for the deformed nuclei in the middle of the rare earth and actinide regions, these spin values are within experimental reach today.

* Invited talk presented at the "High Spin Physics 2001" NATO Advanced Research Workshop, dedicated to the memory of Zdzisław Szymański, Warsaw, Poland, February 6–10, 2001

It appears that nuclei with only a few particles outside closed shells show especially interesting features [1]. Thus, the maximum spin state in the closed shell configuration is generally favoured energetically. Higher spin configurations are then formed when particles are excited across the gaps. Terminating bands of this kind have been studied e.g. in nuclei with a few valence particles outside $^{56}$Ni, $^{100}$Sn and $^{146}$Gd. We will concentrate on two nuclei of this kind, namely $^{62}$Zn [2] and $^{154}$Dy [3], where particle-hole excitations across the gap and within the valence shells lead to configurations with higher and higher maximum spin values. The corresponding terminating bands are low in energy, which has made it possible to observe them in experiment over large ranges in spin.

When classifying terminating bands, it is instructive to consider their behaviour close to the maximum spin values, i.e. close to termination [4, 1]. Detailed calculations and comparisons with experiment show that their energy vs. spin curves, $E(I)$, are strongly dependent on configuration, i.e. how the valence particles are distributed over the $j$-shells. These properties can be used to get reliable configuration assignments for many high-spin rotational bands.

## 2. Calculational procedure - configurations for $Z = 30$.

The low-energy configurations of $^{62}$Zn are illustrated in Fig. 1. In the upper panel, the single-proton orbitals are shown as a function of rotational frequency, $\omega$, at a constant deformation, $\varepsilon_2 = 0.30, \gamma = 20$ and $\varepsilon_4 = 0$. The origin of the orbitals at spherical shape is traced schematically on the left. This left part is thus essentially a standard Nilsson diagram [5], but drawn at 'constant triaxiality', $\gamma = 20°$. A Nilsson (modified oscillator) potential is used including $\vec{l} \cdot \vec{s}$- and $l^2$-terms to put the spherical subshells at their proper places, $\varepsilon_2, \gamma$ and $\varepsilon_4$ deformations and a cranking term, '$-\omega j_x$'.

In the orbitals shown as functions of rotational frequency in Fig. 1, so called virtual crossings have been removed [6] so that diabatic orbitals are formed. These diabatic orbitals behave smoothly as functions of rotational frequency (and also as functions of deformation). The diagonalization is carried out in the rotating harmonic oscillator basis, see e.g. [7], which means that the most important deformation degrees of freedom, $\varepsilon_2$ and $\gamma$, as well as the rotation is included in the basis and thus treated exactly. Furthermore, the $N$-shells (the $N_{rot}$-shells) of this basis are treated as pure [6], i.e. the (small) couplings between them, due to $\varepsilon_4$-deformations and and $\vec{l} \cdot \vec{s}$- and $l^2$-terms, is neglected. Noting that also signature $\alpha$ is a preserved quantum number, we can put the labels $(N, \alpha)$ on the orbitals as shown in Fig. 1 (where black lines are used for $\alpha = 1/2$ and grey wide lines for $\alpha = -1/2$).

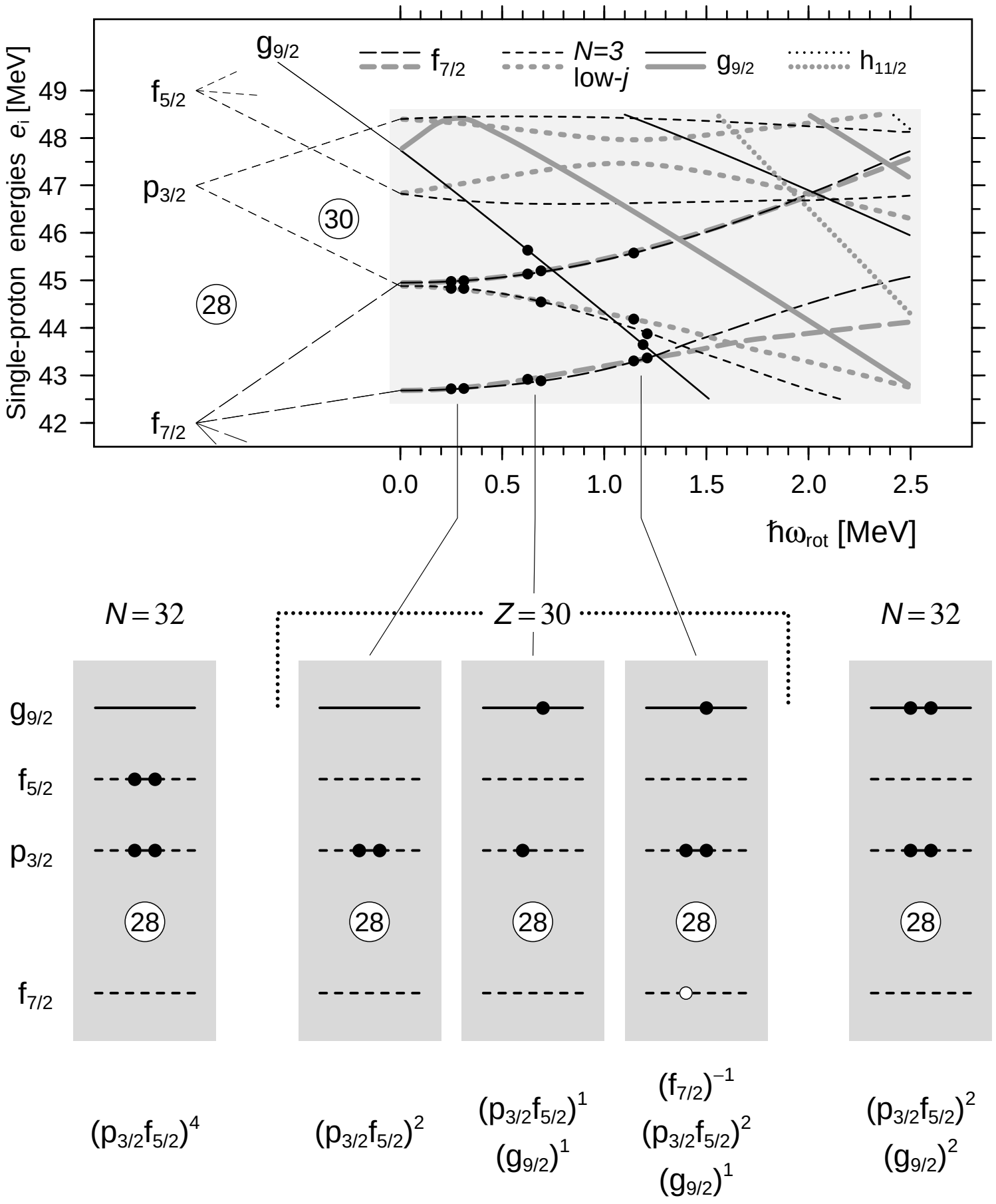

Fig. 1. In the upper panel, the diabatic single-particle orbitals are shown as functions of quadrupole deformation $\varepsilon_2$ for constant triaxiality, $\gamma = 20°$ to the left and for constant deformation as functions of rotational frequency to the right. Different line types are used to indicate how the orbitals are separated into different groups used to define configurations. The filling of these orbitals in the most important $Z = 30$ configurations is illustrated. In the lower panels, the three configurations are illustrated in a more compact way, where the signature of the valence particles is not indicated. Furthermore, the $N = 32$ configurations of the low-$I$ and high-$I$ bands in $^{62}$Zn are shown on the left and on the right, respectively. Note that the different $j$-shells refer to 'deformed orbitals' having their main components in these $j$-shells and that we do not distinguish between e.g. $p_{3/2}$ and $f_{5/2}$ orbitals; both of these subshells belong to the group 'low-$j$ $N = 3$ orbitals'.

A further distinction is that the high-$j$ orbitals (the intruder orbitals) are identified [8, 9]. Practically, the division into high- and low-$j$ orbitals is achieved by calculating the expectation value of $j^2$ at some low frequency, $\omega \approx 0.3$ MeV, and then select those orbitals having largest value of $\langle j^2 \rangle$ as high-$j$ orbitals, following them for all $\omega$-values. Thus, in the $N = 3$ shell, the four levels with highest $\langle j^2 \rangle$ are selected as $f_{7/2}$ orbitals while the other six orbitals have a mixed $(p_{3/2}f_{5/2}p_{1/2})$ character, or maybe rather a $(p_{3/2}f_{5/2})$ character for those orbitals shown in the figure because the high-lying $p_{1/2}$ subshell should have only small amplitudes in these orbitals. Note also that all the $N = 4$ orbitals in Fig. 1 are high-$j$ orbitals so they are denoted $g_{9/2}$ and similarly, the $N = 5$ orbitals are denoted $h_{11/2}$.

Once the orbitals have been classified, it is straightforward to define configurations as is also illustrated in Fig. 1. For 30 protons, the lowest energy is obtained if the orbitals below the spherical $Z = 28$ gap are filled and the two remaining particles are placed in the low-$j$, $N = 3$ orbitals (the $(p_{3/2}f_{5/2})$-orbitals) in Fig. 1. It is evident that at the deformation chosen in Fig. 1, this is the lowest energy configuration at low spin, i.e. at small frequencies. In the schematic drawing in the lower part of Fig. 1, the configuration with all orbitals below $Z = 28$ filled (the $f_{7/2}$ subshell filled) is taken as reference and the two additional particles are then placed in orbitals above the gap, $p_{3/2}$. Note that in our labeling, we make no distinction between the $p_{3/2}$ and $f_{5/2}$ subshells, so it had been equivalent to place both of these particles in $f_{5/2}$ or one of them in $p_{3/2}$ and the other in $f_{5/2}$. In the schematic figure, the signatures are not indicated but it is evident (see upper panel) that the lowest energy is obtained if the two valence particles have different signatures. Another important quantity is the maximum spin value for two particles in $(p_{3/2}f_{5/2})$-orbitals which is easily obtained as $5/2+3/2=4\hbar$.

The next question is then which excited configurations can be defined. One possibility is to put both $(p_{3/2}f_{5/2})$-particles in the same signature but that is not very interesting leading to a higher energy but not bringing in more spin. More interesting is to lift one particle from either the $f_{7/2}$ or $(p_{3/2}f_{5/2})$ orbitals to $g_{9/2}$. Indeed, at the deformation of the figure, this particle would rather be excited from $f_{7/2}$ but at a smaller deformation, all $f_{7/2}$ orbitals are calculated below the $(p_{3/2}f_{5/2})$ orbitals. The configuration with one particle excited from $(p_{3/2}f_{5/2})$ to $g_{9/2}$ is thus illustrated in the middle of Fig. 1. In this case, signature $\alpha = 1/2$ is strongly favoured for the $g_{9/2}$ particle while the signature splitting is smaller for the $(p_{3/2}f_{5/2})$ orbital so that, at this deformation, it is reasonable to consider both these signatures leading to two bands at similar energies but with different total signatures (even and odd spins). Depending on the signature of the $(p_{3/2}f_{5/2})$ particle, the maximum spin is either $9/2+5/2=7\hbar$ or $9/2+3/2=6\hbar$. If we want to make even higher spins for 30 protons, we naturally excite one particle from $f_{7/2}$ to $(p_{3/2}f_{5/2})$ as illustrated to the left in Fig. 1. Note that in this case, the two signatures of the $f_{7/2}$ hole are very close to degenerate so that two degenerate configurations are formed with different total signature (even or odd spin) depending on if the hole is in one or the other signature. If we want to consider even higher excitations for 30 particles, the next natural step is to excite another particle from $f_{7/2}$ to $g_{9/2}$ but in that case, combined with some similar neutron configuration, the equilibrium deformation becomes even larger, so that a superdeformed band is formed.

Fig. 1 is drawn for protons but it is straightforward to define configurations for neutrons in an analogous manner. Furthermore, the deformation is chosen as illustration and a similar procedure is carried through for all other deformations in a mesh in the $(\varepsilon_2, \gamma, \varepsilon_4)$-space. Consequently, configurations can be defined consistently in the full deformation space. The total energy, $E \equiv E_{tot}$, is now obtained as the sum of the single-particle energies but with a renormalization [10, 11] to make sure that the average dependence on deformation and on angular momentum is correct

$$E_{tot} = \sum_{occ} e_\nu + \left( E_{RLD} - \left\langle \sum_{occ} e_\nu \right\rangle \right), \tag{1}$$

i.e. the average sum $\langle \sum_{occ} e_\nu \rangle$ is replaced by the rotating liquid drop energy $E_{RLD}$. The total spin is obtained with no renormalization as the sum of the expectation values of $j_x$

$$I = \sum_{occ} \langle j_x \rangle . \tag{2}$$

Energy vs. spin curves can now be calculated for the fixed configurations at the different mesh points [6]. It thus becomes possible to draw total energy surfaces at fixed spins and find the energy minimum in the $(\varepsilon_2, \gamma, \varepsilon_4)$-space for each configuration. We thus obtain the total energy $E$ and equilibrium deformation as functions of the spin $I$. The energies can be compared directly with experiment.

Let us also point out that for a general understanding of the evolution of rotational bands, we have found the rotating harmonic oscillator model very useful. Thus, introducing some approximations which can be well motivated at small deformations, Cerkaski and Szymański [12] have developed analytical expressions for energies and corresponding deformations, illustrating how rotational bands evolve towards termination. These expressions are repeated in ref. [13, 1]. Numerical calculations based on the exact analytical expressions [14] for the single-particle energies have been published in ref. [15].

## 3. Rotational bands terminating from $I = 10$ to $I = 24$ in $^{62}$Zn.

Let us now consider a specific $Z = 30$ nucleus, namely $^{62}$Zn with 32 neutrons [2]. The lowest energy neutron configuration will then correspond to a closed $N = 28$ core and the 4 valence neutrons in the low-$j$, $N = 3$ orbitals, $(p_{3/2}f_{5/2})$ as appears natural from the upper panel of Fig. 1 and illustrated in the lower left panel. The calculated energies for this configuration are shown in the right panel of Fig. 2 where they are compared with the observed energies shown in the left panel [2].

It is easy to calculate the maximum spin of this configuration as $4+6 = 10\hbar$ which is also the highest spin state in the observed band. In the lowest energy excited configuration which goes to higher spin values than $I = 10$, one neutron is excited from the second $(p_{3/2}f_{5/2})$, $\alpha = -1/2$ orbital to the lowest $\alpha = 1/2$, $g_{9/2}$ orbital as would be expected from the the single-particle diagram, see Fig. 1. In this configuration, the highest spin is $13\hbar$ which is consistent with experiment. In the next excited configuration, one proton is also excited, see middle lower panel of Fig. 1, leading to a total configuration with a maximum spin of $16\hbar$. If one more neutron is excited to $g_{9/2}$, the configuration shown in the lower right panel of Fig. 1 is obtained, and the maximum spin increases to $19\hbar$. Even higher spin states are most easily formed if a proton is excited across the $Z = 28$ gap resulting in the third proton configuration of Fig. 1. Combining it with the configurations with 1 and 2 $g_{9/2}$ neutrons, the two highest spin configurations ($I_{max} = 21\hbar$ and $I_{max} = 24\hbar$) in Fig. 2 are obtained. For these bands, both signatures are about equally favoured because of the signature degeneracy of the highest $f_{7/2}$ orbital.

It is evident that except for the very low-spin states, there is an impressive agreement between calculations and experiment in Fig. 2. A special feature is that the transition energies close to termination increase strongly with spin in some bands while they rather decrease in other bands. Such features are seen more clearly if the energies are shown relative to an average rotational energy, e.g. the energy for rigid rotation, $(\hbar^2/2\mathcal{J}_{rig})I(I+1)$, where $\mathcal{J}_{rig}$ is calculated at a small deformation, see left panel of Fig. 3. In that figure, the valence space configurations of Fig. 2 are shown together with one signature of the core-excited configurations. The different transition energies close to termination means that some bands are essentially constant in this plot while other slope upwards or downwards, or first downwards at lower spin values and then upwards before termination. In the schematic bands shown to the right in Fig. 3, we introduce a labelling for the different types of bands depending on how the last spin units before termination are obtained, favoured, unfavoured or 'rigid rotation like' termination [1]. Note that these labels only refer to the relative energies in the bands, not to their absolute excitation energy. The band showing unfavoured termination is drawn as decreasing at intermediate spin values, then showing a minimum and finally increasing before it terminates. This is how most observed bands showing unfavoured termination behave (if not, the termination would gen-erally occur high above yrast), but bands of this type could also increase for all spin values when drawn as in Fig. 3.

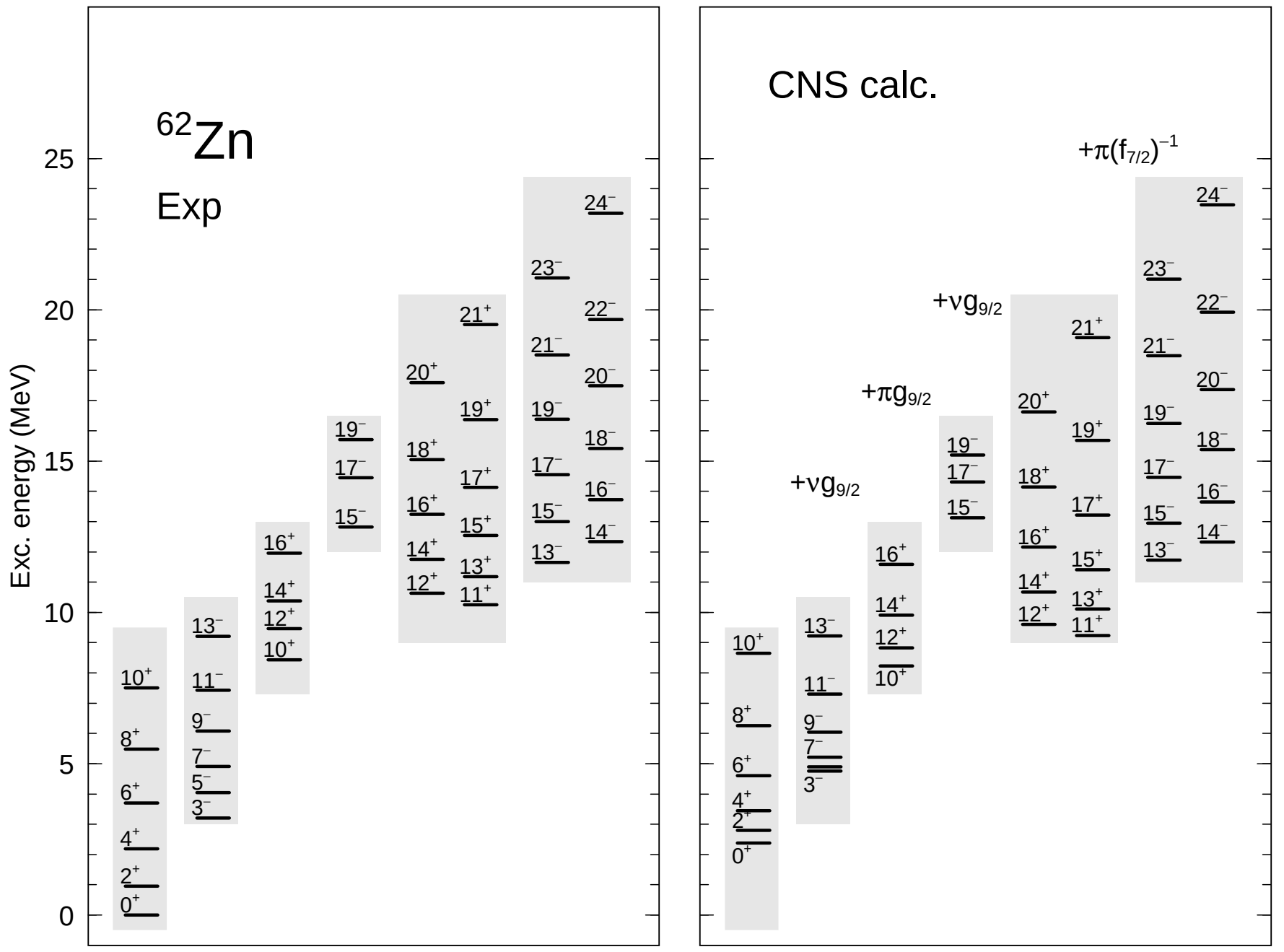

Fig. 2. The low-lying observed bands in $^{62}$Zn are compared with the calculated bands assigned to them, all shown to their maximum spin values.. In the ground band, the 2 valence protons and the 4 valence neutrons occupy low-$j$ $N = 3$ orbitals (orbitals of ($p_{3/2}$, $f_{5/2}$) character) as illustrated in Fig. 1. The next three bands are formed from successive excitations of a proton or a neutron to the $g_{9/2}$ orbitals as indicated in the right panel. In these bands, only the favoured signature is shown. The proton configuration of the bands with maximum spin $16^+$ and $19^-$ is thus the one shown in the middle in Fig. 1. With the right-most proton configuration instead, the maximum spin increases by $5\hbar$ resulting in the bands with maximum spin $21^+$ and $24^-$, respectively. Because the two signatures are degenerate for the $f_{7/2}$ hole (except close to termination), both even (signature $\alpha = 0$) and odd spin ($\alpha = 1$) states are drawn for these bands. Note that some observed bands tend to get larger energy differences close to termination while other bands show the opposite feature and that these tendencies are reproduced in the calculations where standard parameters from ref. [6] are used.

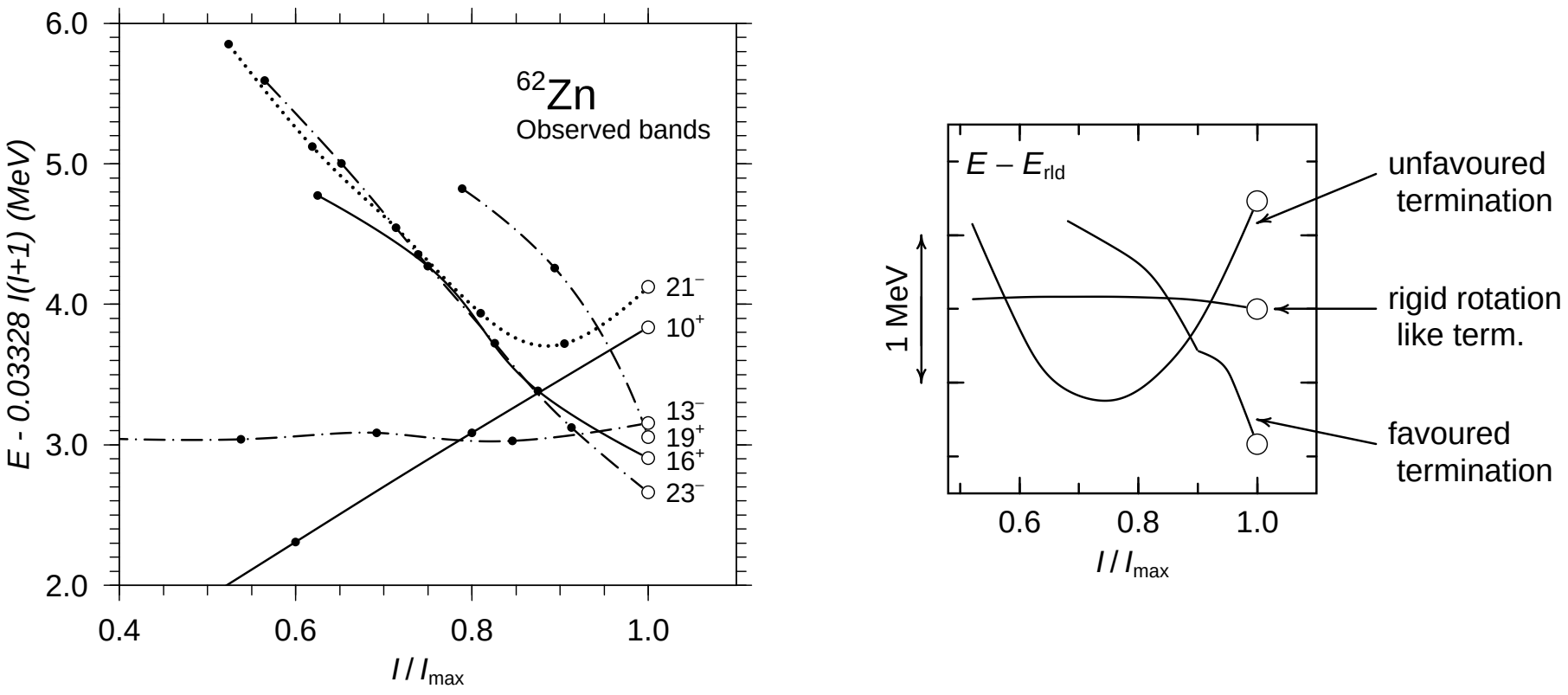

Fig. 3. In the left panel, observed terminating bands of $^{62}$Zn are drawn relative to a rigid rotor reference with the spin measured in units of the maximum spin, $I_{max}$, in the respective configurations (left panel). In the right panel, typical behaviours of terminating bands as they show up in this type of diagram are illustrated and a labelling is introduced.

When comparing the observed and schematic bands in Fig. 3, it is evident that all three types of terminations are observed in $^{62}$Zn. Thus, the ground band terminating at $I = 10^+$ and the band terminating at $21^-$ are examples of unfavoured terminations and the terminations at $13^-$ and $19^+$ are ideal examples of 'rigid rotation like' and favoured terminations, respectively. These bands illustrate some general rules, namely that a few particles distributed over several high-$j$ shells correspond to a low energy cost when building angular momentum and that the energy cost increases if the shells start to be close to half-filled, exemplified e.g. by the $(p_{3/2}f_{5/2})$ neutrons in the ground band. The ground band also illustrates that if many particles occupy low-$j$ shells, the energy cost will be high. Alternatively, these rules can be summarized by stating that the energy cost per angular momentum unit will be low if the average contribution from each particle is large while the energy cost becomes higher if the average contribution to the angular momentum is smaller. This is clearly seen, e.g. by comparing the bands terminating at $10^+$ and $19^-$ where in both cases, 6 valence particles are active.

A further observation is that particles and holes tend to counteract each other which could be understood from their different shape driving properties. Thus, a high-$j$ particle contributing with its full angular momentum will essentially rotate in a circular orbital around the equator of the nucleus (the spherical core) leading to a total mass distribution which is oblate while a corresponding hole will dig a 'hole' around the equator so that the total mass distribution tends to become prolate. Therefore, in a terminating state where both particles and holes contribute to the angular momentum, they will have counteracting shape driving properties making this state less favoured energetically. This is seen e.g. when comparing the bands terminating at $16^+$ and $21^-$ in Fig. 3 which differ by a particle-hole excitation $(f_{7/2})^{-1}(p_{3/2}, f_{5/2})^1$. It is evident that the band with no $f_{7/2}$ hole shows a much more favoured termination than the band with a hole.

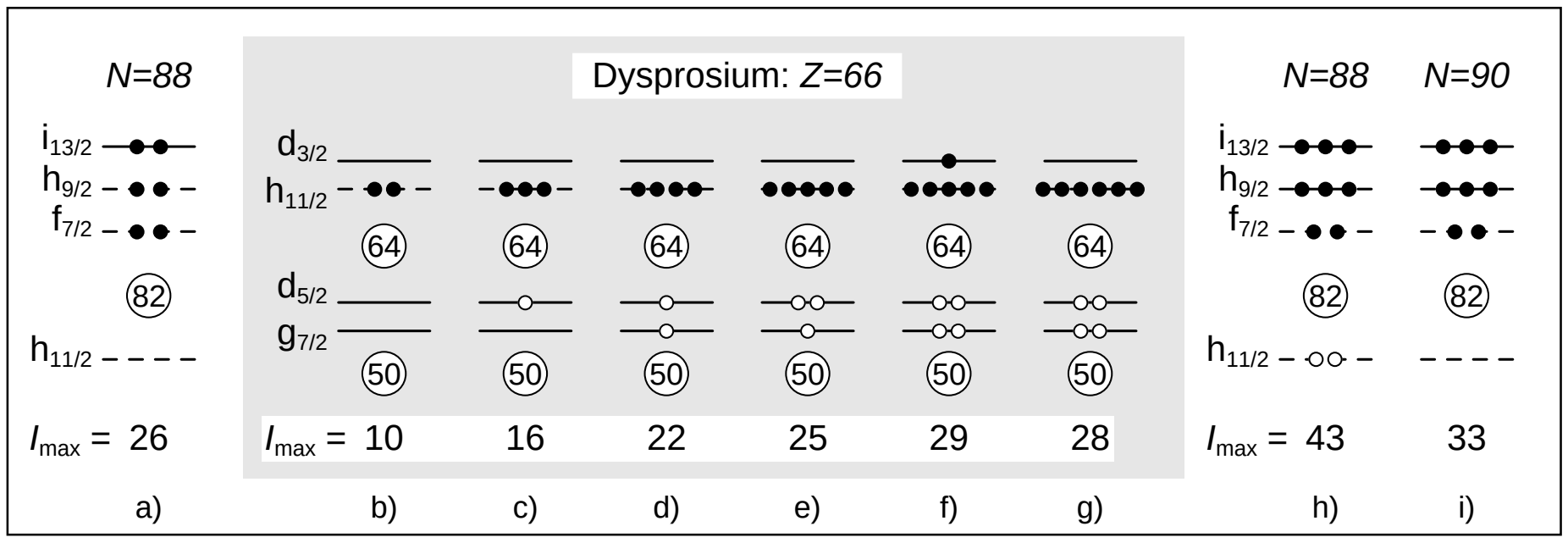

Fig. 4. Illustration of the $Z = 66$ configurations which appear to build terminating bands in $^{154}$Dy and $^{156}$Dy. These configurations are shown in between the two $N = 88$ configurations which are dominating when build terminating bands and collective bands, respectively in $^{154}$Dy. The $N = 90$ configuration which is active when building the observed rotational bands in $^{156}$Dy which terminate for $I_{max} \approx$ 60 is shown to the right. The maximum spin value in the different configurations is also shown.

## 4. Particle-hole excitations across the semi-magic $Z = 64$ shell

Turning to heavier nuclei, another interesting particle number is $Z = 66$ with two particles outside the $Z = 64$ subshell. The corresponding closed shell configuration as well as configurations with particles excited across the shell gap are illustrated in Fig. 4, where also the maximum proton spin values in the respective cases are given. All of these proton configurations have been observed to termination or close to termination, either in the $N = 88$ nucleus $^{154}$Dy [3] or in the $N = 156$ nucleus $^{156}$Dy [16]. The relevant $N = 88$ and $N = 90$ configurations are shown to the left and right in this figure. In a similar way as for the $f_{7/2}$ protons in the $N = 3$ shell, the high-$j$ $h_{11/2}$ neutrons are distinguished from the other $N = 5$ neutrons for these configurations with a few neutrons outside the $N = 82$ core. For the protons, an even more detailed division has been made [17] so that labels are put on the different 'pseudospin' groups in the $N = 4$ shell, $g_{9/2}$, $(g_{7/2}d_{5/2})$ and $(d_{3/2}s_{1/2})$, respectively. Especially, we keep track of the number of particles excited from the orbitals of $(g_{7/2}d_{5/2})$ character to orbitals of $(d_{3/2}s_{1/2})$ character also in regions where they mix in energy. The two proton configurations e) and f) in Fig. 4 differ by such an excitation.

Let us now consider some selected observed bands in $^{154}$Dy as shown in the upper panel of Fig. 5, namely the bands interpreted as built from the proton configurations b) - d) in Fig. 4 combined with the $N = 88$ configuration illustrated to the left in that figure. Furthermore, one collective band is shown, interpreted as built from the proton configuration g) in Fig. 4 combined with the neutron configuration h). It is mainly the two neutron holes in the $h_{11/2}$ orbitals which give this band its special character being almost constant when drawn vs. a rigid rotation reference as in Fig. 5, i.e. the energy in the $I = 25 - 50$ range can approximately be expressed as $E = E_0 + const. \cdot I(I+1)$. This suggests that this band is strongly collective in this spin range in agreement with calculations (see lower panel of Fig. 5) where, however, the band can be followed to termination at much higher spin, $I = 71\ \hbar$. This terminating state is calculated high above yrast and probably impossible to observe in experiment.

Coming back to the other observed in Fig. 5, they are all down-sloping in the spin range $I = 30-40$. Furthermore, they are observed to the maximum spin values when combining the neutron and proton configurations in Fig. 4. These facts together with the general agreement with calculations, see lower panel of Fig. 5, show convincingly that we have found the correct interpretation of these bands. This interpretation is also supported by lifetime measurements in the positive parity bands observed to $I^\pi = 36^+$ and $I^\pi = 48^+$ and by the absence of alternative interpretations, at least for the bands observed to $I^\pi = 41^-, 42^-$ and $I^\pi = 48^+$.

It is thus evident that the $Z = 66$ configurations of Fig. 4 with up to two protons excited across the $Z = 64$ gap have been observed in experiment while there is no experimental counterpart of the configurations with 3 or 4 protons excited. These configurations are however observed in the $N = 90$ isotope $^{156}$Dy [16], in combination with the neutron configuration in Fig. 4i) with 8 valence neutrons in the $(h_{9/2}f_{7/2})$ and $i_{13/2}$ orbitals . They are well described by calculations [16, 17] to spin values close to termination, $I \approx 60\hbar$. Calculations for the corresponding configurations in $^{154}$Dy are presented in Fig. 5. Because they are yrast or close to yrast, it should be possible to identify them in experiment; i.e. their observation is a challenge for future experiments.

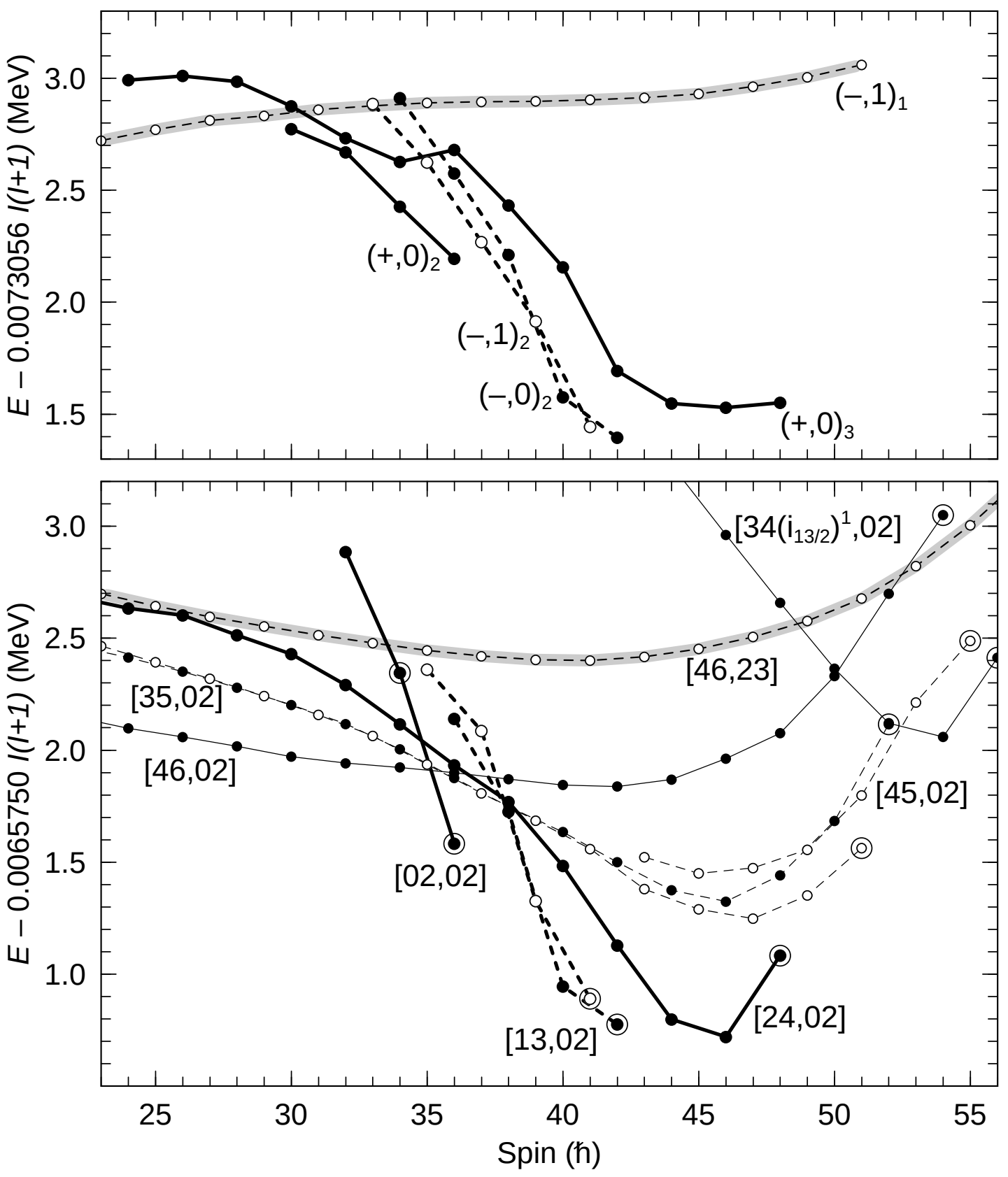

Fig. 5. The rotational bands clearly observed to termination in $^{154}$Dy are shown in the upper panel together with the collective band identified to highest spin values. Calculated configurations assigned to these bands are shown in the lower panel using thick lines and a grey shading, respectively. Furthermore, some bands calculated low in energy and terminating for $I_{max} \approx 50 - 55\hbar$ are shown with thin lines. The calculated configurations are labeled as $[p_1p_2, n_1n_2]$ where $p_1$ is the number of proton holes of $(g_{7/2}d_{5/2})$ character, $p_2$ the number of $h_{11/2}$ protons, $n_1$ is the number of $h_{11/2}$ neutron holes and $n_2$ is the number of $i_{13/2}$ neutrons. The number of $(d_{3/2}s_{1/2})$ protons and $(h_{9/2}f_{7/2})$ neutrons is not given explicitly but determined from the total number of protons and neutrons. At high spin values, configurations with one proton excited to the lowest $i_{13/2}$ orbital become competitive in energy as illustrated for the configuration terminating at $I_{max} = 56\hbar$. In these calculations for $^{154}$Dy, we have used the standard parameters for this region [18], previously applied e.g. to $^{158}$Er [19] and $^{156}$Dy [16].

It is also interesting to study the variation of the energy functions close to termination in these $^{154}$Dy bands. When drawn as in Fig. 5, the bands are steeply down-sloping for the configurations with none or one proton excited across the gap, then becoming softer with two particles excited and turning into regular smooth terminating bands with 3 or 4 particles excited across the gap. This is in agreement with the rules discussed above in connection with $^{62}$Zn. For example, the holes in the core tend to make the termination less favoured energetically and the fact that the $h_{11/2}$ shell becomes close to half-filled has the same effect.

## 5. 'Magnetic bands' in configurations with one ($g_{7/2}d_{5/2}$) hole in the $Z = 64$ core.

With one hole in the $Z = 64$ core, two bands which are almost signature degenerate and connected by $M1$ transitions are formed. Indeed, $^{154}$Dy is a very special case where such a 'magnetic band' is observed [3] to termination at very high spin for both signatures, $I = 41$ and $I = 42$. In this band, all $M1$-transitions down to $I = 33$ have been observed while it has not been possible to observe the $E2$-transition for the decay of the $42^-$ state to the $40^-$ state, illustrating the relative strength of the $M1$'s relative to the $E2$'s. The present principal axis cranking calculations reproduce these features reasonably well even though there are important discrepancies when it comes to the detailed comparison of the transition energies. One problem is that oblate states with the same distribution of particles over the $j$-shells (same configuration) but with one neutron anti-aligned start to become lower in energy for spin values $6\hbar$ below the terminating spin values. This means that it is not possible to follow the terminating bands to lower spin values in the present formalism.

In early calculations [4] for $^{158}$Er, terminating bands connected by $M1$-transitions were predicted. These bands have configurations analogous to those of $^{154}$Dy discussed above, but with two additional valence protons as well as two valence neutrons. This leads to terminating spin values around or slightly above $I = 50$. We have repeated the calculations of ref. [4] using our present formalism where we can keep track of the number of particles excited from orbitals of ($g_{7/2}d_{5/2}$) character to orbitals of ($d_{3/2}s_{1/2}$) character. The result is presented in Fig. 6 supporting previous calculations where some states were interpolated assuming no signature splitting. These bands in $^{158}$Er are predicted to have almost constant $M1$ transition energies in a large spin range up to termination . Because of the larger number of valence particles (higher spins), the average of these transition energies are however calculated as $0.50 - 0.55$ MeV in $^{158}$Er, which can be compared with the observed values around 0.35 MeV in $^{154}$Dy. These larger transition energies along the terminating bands means that oblate states with one spin vector anti-aligned become less competitive in $^{158}$Er so that the terminating bands can be followed in a larger spin range. These bands in $^{158}$Er are calculated as yrast and very close to yrast. Thus, there should be good possibilities to observe them in future experiments even though it should also be noted that a number of other bands are calculated at similar energies.

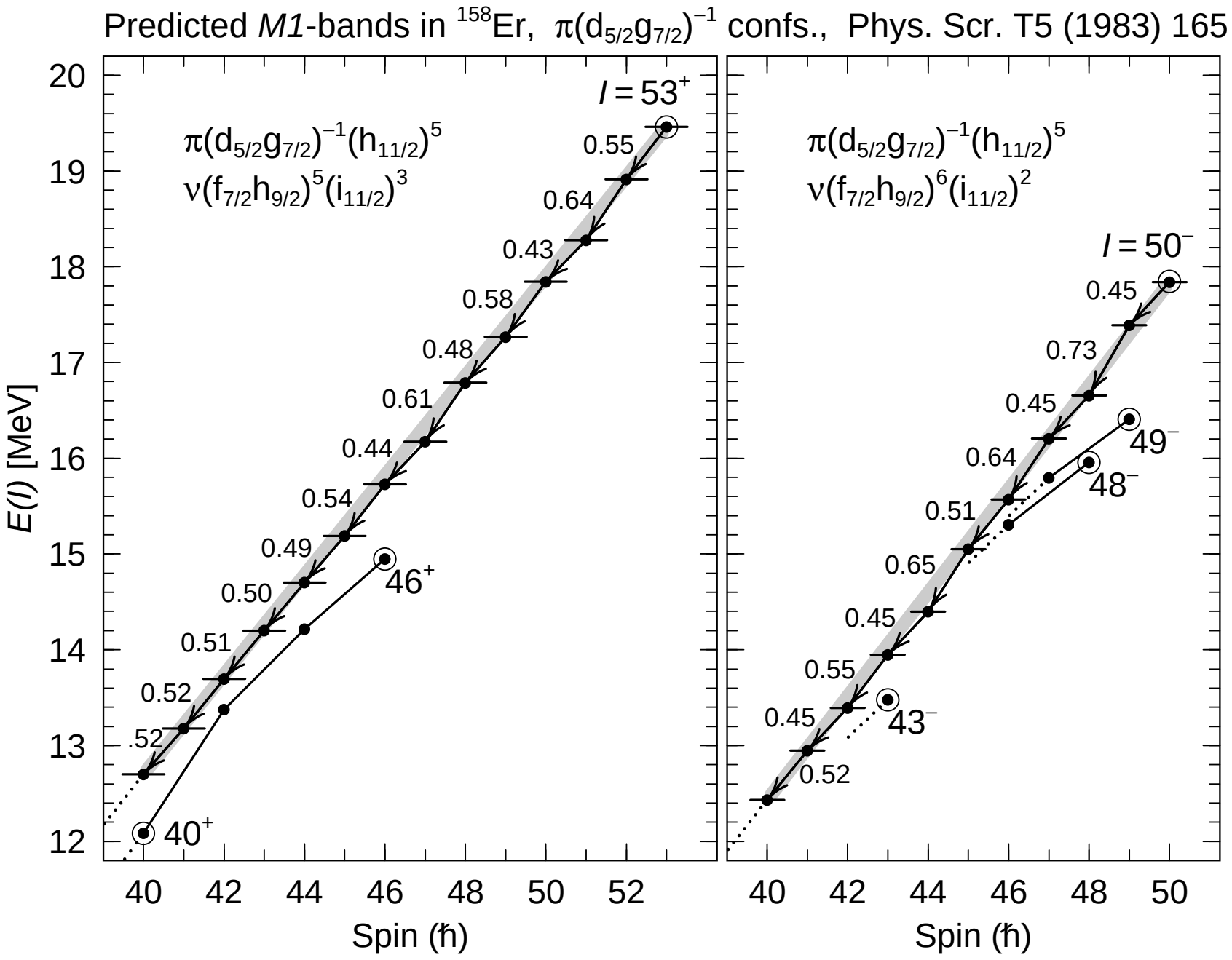

Fig. 6. Predicted $M1$-bands in $^{158}$Er in configurations with one proton hole in the $(g_{7/2}d_{5/2})$ orbitals. Compared with the similar plots presented in ref. [4], the single-particle parameters fitted to this region [18] have been used in the present calculations together with a formalism [9, 17] where configurations are specified in more detail. The two configurations should be labelled as [15,03] and [15,02] according to the conventions used in Fig. 5. In addition to these two bands, a third [15,03] band with a different signature of the 5 $(g_{7/2}d_{5/2})$ neutrons and terminating at $I = 52^-$ is calculated at similar excitation energies. The calculated terminating states at $I^\pi = 46^+, 43^-, 48^-$ and $49^-$ are shown for comparison and the corresponding bands are sketched. Apart from these states, which are known to be yrast, no other bands are calculated clearly below the $M1$-bands but several bands are calculated at similar energies. The calculated energies of the $M1$ transition are given explicitly (in MeV), but it is questionable if the small deviations from constant transition energies (indicated by the thick shaded lines) are numerically significant.

## 6. Summary

We have analyzed nuclear high-spin states using the configuration dependent cranked Nilsson-Strutinsky model with the Nilsson potential. Our method [6, 1] is based on single-particle diagonalization in a rotating harmonic oscillator potential, which makes it straightforward to introduce a configuration definition which can be used for all deformations and coupling schemes. It thus becomes possible to follow the corresponding rotational bands as functions of the nuclear spin. This is especially important when the nuclear spin is close to its maximal value $I_{max}$, i.e. close to termination, where large shape changes occur.

Main emphasis was put on the Zn and Dy isotopes, especially ${}^{62}_{30}\mathrm{Zn}_{32}$ and ${}^{154}_{66}\mathrm{Dy}_{88}$. The high-spin configurations in these nuclei can be characterized by the number of protons excited across the shell gaps at $Z = 28$ and $Z = 64$, respectively. Several rotational bands based on closed core configurations are followed to termination for $I_{max} = 10 - 19$ in ${}^{62}$Zn. These bands show characteristic features depending on the distribution of the valence particles over the orbitals of low-$j$ character ($\mathrm{p}_{3/2}\mathrm{f}_{5/2}$) and high-$j$ character ($\mathrm{g}_{9/2}$), respectively. With one proton excited across the $Z = 28$ shell gap, smooth terminating bands are formed which are observed in the approximate spin range $[I_{max}/2,\ I_{max}]$ with $I_{max} = 20 - 24$. These bands decay by stretched $M1$ and $E2$ transitions of similar strength.

Terminating bands of different character are observed in the closed core and in the 1p-1h and 2p-2h configurations of ${}^{154}$Dy. Calculations and comparisons with ${}^{156}$Dy suggest that it should be possible to observe bands to termination also in the 3p-3h and 4p-4h configurations These terminating states become less favoured energetically with increasing number of particle- hole excitations. In analogy with ${}^{62}$Zn, stretched $M1$-transitions are observed in the 1p-1h configuration of ${}^{154}$Dy where the terminating spin values for the two signatures are $41^-$ and $42^-$ and where the $E2$ collectivity is so small that it has not been possible to observe one of the $E2$-transitions. The prediction of similar 'magnetic bands' terminating at even higher spin values in ${}^{158}$Er ($I_{max} \geq 50\hbar$) was discussed.

Financial support from the Swedish Natural Science Research Council is gratefully acknowledged. Work at Mississippi State University is supported by US DOE grant DE-FG02-95ER40939.